\documentclass[amsmath,amssymb,aps]{revtex4} 
\usepackage{graphicx}
\usepackage{epstopdf}
\usepackage{amsmath}
\usepackage{bm}

\begin{document}

\title{General Relativistic Quantum Mechanics \\ deriving \\ Electroweak and Gravitational Interactions}

\author{Kimihide Nishimura}
\email{kimihiden@dune.ocn.ne.jp}
\affiliation{Nihon-Uniform, 1-4-1 Juso-Motoimazato, Yodogawa-ku Osaka 532-0028 Japan}
\date{\today}

\begin{abstract}
A gauge theory with an indefinite metric without negative probabilities is given by extending quantum mechanics, 
where a general metric is introduced, and the invariance under the general linear transformation is imposed on the space of quantum states. 
On this basis, we construct and investigate a chiral sextet model, which has one more Lorentz symmetry in the gauge space, to derive much properties of the standard electroweak theory, and also Einstein gravity, when the double Lorentz symmetry spontaneously fuses into one.
\end{abstract}

\maketitle
\section{Introduction}
One of the hindrances to develop unified understandings of elementary particles and forces, like an SU(5) theory\cite{GG},  seems to reside in the inability of gauge theories to include Lorentz symmetry in the gauge group, 
due to which we have been not yet able to investigate properly the phenomenon of spontaneous spacetime symmetry breakdown \cite{Bjorken,Eguchi,BK,KN1,KN2,KN3}. 
This obstacle can be removed, however, by generalizing quantum mechanics by introducing the general metric in the space of quantum states and, as in the general relativity theory, by requiring the invariance under general linear transformations. 

Historically, though it seems that Dirac had been aware of a possibility to extend relativistic quantum mechanics further to include general relativity, 
he did not do that by saying ``gravitational effects are negligible in atomic theory"\cite{Dirac1942}.  
Instead, he tried to solve divergence problems in QED by reviving an old quantum theory with negative energies and negative probabilities, which had been already removed by the method of ``second quantization". 
Pauli, however, had been aware that Dirac's approach concerned with the indefinite metric in the space of quantum states, which makes quantum mechanics invariant under general linear transformations\cite{Pauli}. 
It is this viewpoint that provides the germ of the ``general relativistic quantum mechanics". 

We show that the generalized quantum mechanics alters the definition of probability, 
and requires the conservation of the quantum metric. 
No negative probabilities appear after the second quantization.

As the result, the general quantum mechanics enables us to have a consistent gauge theory including Lorentz symmetry in its subgroup, and to investigate spacetime symmetry breakdown in the gauge space for unifying the elementary particles and forces.

An interesting possibility arising there is that the generalized gauge theory can have double Lorentz symmetry; one is in the gauge space, and the other is in the real spacetime. 
Further, there appears a particular kind of symmetry breaking, in which the overlapping Lorentz symmetries fuse into one. In this case, the same spacetime symmetry survives in the effective theory as that broken,  
which solves the long-standing dilemma in the phenomenological aspect of the Lorentz violation. 

We apply this fusion mechanism to the chiral sextet model for obtaining the electroweak and gravitational interactions. 
In particular, the theory predicts the Weinberg angle: $\sin^2\theta_W=1/4$, 
the Higgs boson mass: $m_H=\sqrt{2}m_Z$, 
and the quartic coupling constant for the Higgs potential: $\lambda=g_W^2/3$. 
The Higgs doublet, the right-handed lepton, and the gravitational equations are also derived from the model, when the complex six dimensional local Lorentz group breaks down to the real four dimensional Lorentz group and an SU(2)$\times$U(1).

\section{general relativity in quantum mechanics}
The space of quantum states is a complex linear projective space. 
It is projective, since, due to the quantum mechanical superposition principle\cite{Dirac}, 
a wave function, or a state vector, $\Psi$ multiplied by a non-zero number represents the same quantum state. 
However, the probability, or the norm of the state vector, $\Psi^\dagger\Psi$ defined in the conventional quantum mechanics changes by the scale transformation. 
Then, it will be desirable to improve the definition by introducing a Hermitian metric $\xi$ in the quantum space to make the new probability  
$\Psi^\dagger\xi\Psi$ invariant under the scale transformation.   
Since the phase transformation also leaves the state invariant, we require that the total probability 
\begin{equation}
1=\bar{\Psi}\Psi, \quad \bar{\Psi}:=\Psi^\dagger\xi,
\label{TotalProb}
\end{equation}
should be invariant under the general linear transformation $G$:  
\begin{equation}
\Psi'=G\Psi, \quad \bar{\Psi}'=\bar{\Psi}{G}^{-1}, \quad \xi'=G^{\dagger-1}\xi G^{-1}.
\label{GGT}
\end{equation} 
We suppose here for simplicity and definiteness that a quantum system takes only a finite number of states labeled by an integer ``$i$", and then ask the probability $P_i$ to find the system in state ``$i$". 

In the component representation, Eq.(\ref{TotalProb}) reads
\begin{equation}
1=\sum_{ij}\Psi^{i*}\xi_{ij}\Psi^j=\sum_i\Psi_i^*\Psi^i, \quad \Psi_i:=\sum_j\xi_{ij}\Psi^j.
\label{InvProb}
\end{equation}
Then, $P_i$ will be
\begin{equation}
P_i=\Psi_i^*\Psi^i. 
\end{equation}
The probability in the generalized quantum mechanics is not the product of the wave function and its complex conjugate, 
but the wave function multiplied by the complex conjugate of its reciprocal. 

We observe here the parallelism to the theory of general relativity.  
The invariant probability (\ref{InvProb}) has the parallel with the invariant length squared in general relativity:
\begin{equation}
ds^2=g_{\mu\nu}(x)dx^\mu dx^\nu, 
\end{equation}
where $dx^\mu$ corresponds to $\Psi^i$, 
and the invariant length squared $ds^2$ corresponds to the total probability$=1$. 

We further recognize some similarities and differences from this observation.
Though the components of the wave function correspond to the differentials of the general coordinates, 
no difference exists between the differentials and the coordinates themselves in the linear space of quantum states. 
Accordingly, the metric $\xi$ in the quantum space can not depend on the wave function $\Psi$ itself, 
in contrast to the case of general relativity, where the spacetime metric $g_{\mu\nu}$ depends on the general coordinates.

We next assume the Schr\"{o}dinger equation,
\begin{equation}
i\frac{d\Psi}{dt}=H\Psi,
\end{equation}
where the general metric $\xi$ and the general transformation $G$ are not yet time-dependent.
The general transformation invariance requires that the Hamiltonian transforms as 
\begin{equation}
H'=GHG^{-1}.
\label{GHG}
\end{equation}
The conservation of the invariant probability requires
\begin{equation}
\bar{H}=H, \quad \bar{H}:=\xi^{-1}H^\dagger\xi,
\label{QuasiHermitian}
\end{equation}
since
\begin{equation}
\frac{d}{dt}(\bar{\Psi}\Psi)=i\bar{\Psi}(\bar{H}-H)\Psi=0.
\end{equation}
We call the matrix or the operator satisfying $\bar{Q}=Q$ quasi Hermitian. 

The general metric $\xi_{ij}$ can take the canonical form $\xi_{\alpha\beta}=\eta_{\alpha\beta}$ by the general linear transformation $G$, 
where $\eta_{\alpha\beta}$ is a diagonal matrix, and each of the diagonal elements is $1$ or $-1$. 
Since $\eta^2=1$, we can interpret $\eta$ as a parity operator with eigenvalues $\pm1$, 
whence $\Psi_\alpha=\pm\Psi^\alpha$ results.  

We may call the trace of $\eta$ the signature of $\xi$.  
It is the consequence from the Sylvester's law of inertia that the signature Tr$(\eta)$ is the intrinsic property of $\xi$, 
invariant under the general linear transformations.

Even after fixing $\xi$ in this gauge, 
there still remains the invariance under the special linear transformation $S$, satisfying,
\begin{equation}
S^\dagger\eta S=S\eta S^\dagger=\eta.
\label{SGT}
\end{equation}
If we assign positive integers for the states with positive $\eta$-parity, while negative integers for the negative $\eta$-parity, then the total probability (\ref{InvProb}) reads
\begin{equation}
\bar{\Psi}\Psi=\sum_{\alpha>0}\vert\Psi^\alpha\vert^2-\sum_{\alpha<0}\vert\Psi^\alpha\vert^2.
\end{equation}
Though it conserves for the quasi Hermitian Hamiltonian, the partial probabilities can be negative, greater than one, or divergent, 
which implies that the probability interpretation of wave function will fail. 
However, we rarely consider Hamiltonians which are not Hermitian. 
For a Hermitian Hamiltonian, the condition $\bar{H}=H$ reduces to
\begin{equation}
[H, \xi]=0, \quad{\rm or}\quad [H, \eta]=0.
\label{ConservingGM}
\end{equation}
which implies that the general metric $\xi$ is a constant of motion, or that the $\eta$-parity conserves. 
This condition again has an analogue in general relativity: the invariance of the spacetime metric under the parallel transportation:
\begin{equation}
\nabla_\mu g_{\rho\sigma}=0.
\end{equation} 
Further, when the Hamiltonian is Hermitian, the conventional probability also conserves: 
\begin{equation}
\Psi^\dagger\Psi=\sum_{\alpha>0}\vert\Psi^\alpha\vert^2+\sum_{\alpha<0}\vert\Psi^\alpha\vert^2.
\end{equation}
Then, the positive probability and the negative probability are separately conserved:
\begin{equation}
\sum_{\alpha>0}\vert\Psi^\alpha\vert^2=1, \quad \sum_{\alpha<0}\vert\Psi^\alpha\vert^2=1.
\end{equation}
 Generally, in the canonical gauge, any state $\Psi$ can be decomposed into the eigenstates of $\eta$-parity:
\begin{equation}
\Psi=\Psi_++\Psi_-, \quad \eta\Psi_\pm=\pm\Psi_\pm,\quad \bar{\Psi}_\pm\Psi_\pm=\pm1, 
\quad \bar{\Psi}_+\Psi_-=0. 
\end{equation}
Since $\eta$-parity conserves, the time evolution of the positive wave function always remains positive, and that of the negative wave function always negative. There are no transitions from positive to negative states, and vice versa. 
The domain of the states connected in time evolution is divided into two parts $D_\pm$, in which $D_+$ consists only of positive $\eta$-eigenstates, and $D_-$ only of negative $\eta$-eigenstates. The interactions are unitary in each subspace. 

The dichotomy appearing in the space of quantum states caused by the general metric is also analogous to 
the light-cone in the Minkowski spacetime, separating the spacetime points unattainable by time evolution. 
We next show that the second quantization makes the invariant probability always positive definite. 
\section{second quantization}
We transfer to the many particle system by the method of second quantization.
The Lagrangian for the Schr\"{o}dinger equation is now,
\begin{equation}
L=\bar{\Psi}\left(i\frac{d}{dt}-H\right)\Psi.
\end{equation}
The reality of the Lagrangian requires $\xi=\xi^\dagger$ and  $\bar{H}=H$. 
The canonical momentum for $\Psi^\alpha$ is $\partial L/\partial\dot{\Psi}^\alpha=i\Psi^\dagger_\alpha$, 
and the quantum conditions are
 \begin{equation}
\{\Psi^\alpha,\Psi^\dagger_\beta\}=\delta^\alpha{}_\beta, \quad 
\{\Psi^\alpha,\Psi^\beta\}=0, \quad 
\{\Psi_\alpha^\dagger,\Psi_\beta^\dagger\}=0, 
\label{QuantumCondition}
\end{equation}
where we have assumed that the particles are fermions, and have quantized the wave function $\Psi^\alpha$ by the anti-commutation relations.  We next take $\xi$ in the canonical gauge $\eta$, and express $\eta$ and the special transformation $S$ in the operator form:
\begin{equation}
\eta\Psi^\alpha\eta=\Psi_\alpha, \quad {\Psi'}^\alpha=S\Psi^\alpha{S}^{-1}.
\label{ReciprocalConjugate}
\end{equation} 
Then, owing to the relations (\ref{SGT}), we have
\begin{equation}
{\Psi'}_\alpha^\dagger=S\Psi_\alpha^\dagger{S}^{-1}, 
\end{equation}
which implies that the canonical anti-commutation relations (\ref{QuantumCondition}) are invariant under the special transformation.

Interpreting $\Psi^\alpha$ as an annihilation operator, while $\Psi_\alpha^\dagger$ as a creation operator of a particle, 
a many particle state $\vert\Phi\rangle$ is represented as 
\begin{equation}
\vert\Phi\rangle=(\prod_\alpha\Psi_\alpha)^\dagger\rangle.   
\label{ManyParticleState}
\end{equation}
The standard ket: ``$\rangle$" in Dirac's notation represents the Fock vacuum satisfying $\Psi_\alpha\rangle=0$.  
Then, the special transformations of the state and its dual operator are
\begin{equation}
\vert\Phi'\rangle=(\prod_\alpha{\Psi'}_\alpha)^\dagger\rangle
=(\prod_\alpha S^{\dagger-1}\Psi_\alpha{S}^\dagger)^\dagger\rangle=S\vert\Phi\rangle,
\label{GetaG}
\end{equation}
\begin{equation}
\langle\bar{\Phi}'\vert=\langle\Phi'\vert\eta=\langle\Phi\vert{S}^\dagger\eta=\langle\Phi\vert\eta{S}^{-1}=\langle\bar{\Phi}\vert{S}^{-1},
\end{equation}
where the invariance of the Fock vacuum has been assumed. 
The result shows that the probability: 
\begin{equation}
\langle\bar{\Phi}\vert\Phi\rangle=\langle\Phi\vert\eta\vert\Phi\rangle, 
\label{InvariantProbability}
\end{equation}
is invariant under the special transformations. 
Further, it is positive definite, since
\begin{equation}
\langle\bar{\Phi}\vert=\langle\prod_\alpha\Psi_\alpha\eta=\langle\prod_\alpha\eta\Psi_\alpha\eta
=\langle\prod_\alpha\Psi^\alpha, \quad 
\langle\bar{\Phi}\vert\Phi\rangle=\langle\prod_\alpha\Psi^\alpha\Psi_\alpha^\dagger\rangle=1,
\end{equation}
due to the canonical anti-commutation relations (\ref{QuantumCondition}), 
where the Fock vacuum is assumed invariant also under the operation of $\eta$. 

We remark that the Schr\"{o}dinger equation after the second quantization has the form 
\begin{equation}
i\frac{d}{dt}\vert\Phi\rangle=\hat{H}\vert\Phi\rangle, \quad \hat{H}=\bar{\Psi}H\Psi,
\end{equation}
where the Hamiltonian operator $\hat{H}$ is Hermitian: $\hat{H}^\dagger=\hat{H}$, 
and invariant under the the special transformation : $\hat{H}'=\hat{H}$, 
due to (\ref{GGT}), (\ref{GHG}), and (\ref{QuasiHermitian}), for $\xi=\eta$, and $G=S$.
The invariant probability conserves under the condition:
\begin{equation}
[\hat{H}, \eta]=0.
\label{GMC}
\end{equation}
We have seen that the invariant probability in the second quantization is positive definite, irrespective of the signature of the general metic $\xi$ in the space of quantum states, 
though the consistency of the quantum theory with an indefinite metric without negative probabilities relies on the conservation of the canonical metric (\ref{GMC}). 
We next show that this condition is included in the gauge invariance of the generalized gauge theory 
with an arbitrary spacetime-dependent metric in the gauge space.
  
\section{generalized gauge theory}
The space of quantum states are ordinarily infinite dimensional, where the states are labeled by the coordinates or momenta, spins, charges and the other internal degrees of freedom of particles.
To avoid conflict with spacetime symmetry, we restrict the general transformation invariance only in the gauge space of finite complex dimensions. 
This restriction is equivalent to introduce the spacetime-dependent general metric in the gauge space, and to require the general linear transformation invariance as the gauge symmetry.  
 
We consider for definiteness the Lagrangian density of a left-handed Weyl $d$-plet: $\chi=(\chi^\alpha)$, $(\alpha=0,\cdots d-1)$ in the four dimensional Minkowski spacetime,  
and introduce the gauge field $X_\mu$ to make the theory invariant under the general gauge transformations.
Then, the Lagrangian for $\chi$ is given by
\begin{equation}
{\cal L}_\chi=\bar{\chi}\bar{\sigma}^\mu iD_\mu\chi
=\chi^\dagger\xi\bar{\sigma}^\mu(i\partial_\mu-gX_\mu)\chi
=\chi^\dagger_\alpha\bar{\sigma}^\mu(i\partial_\mu\chi^\alpha-gX_\mu{}^\alpha{}_\beta\chi^\beta),
\end{equation}
which is invariant under the general gauge transformation: 
\begin{equation}
\chi'=G\chi, \quad \xi'=G^{\dagger-1}\xi{G}^{-1}, \quad X_\mu'=GX_\mu G^{-1} +\frac{i}{g}\partial_\mu GG^{-1}.
\label{GTG}
\end{equation}
The reality of ${\cal L}_\chi$ requires the invariance of $\xi$ under the ``parallel transportation": 
\begin{equation}
D_\mu\xi=\partial_\mu\xi+ig\xi(\bar{X}_\mu-X_\mu)=0, 
\quad \bar{X}_\mu:=\xi^{-1}X_\mu^\dagger\xi.
\end{equation}
Owing to the general gauge invariance, $\xi$ can be transformed globally into the canonical metric $\eta$, satisfying $\eta^2=1$, since the space of quantum states is a linear space. 
Then, the gauge field $X_\mu$ becomes quasi Hermitian: $\bar{X}_\mu=X_\mu$, 
or expressed in components: $(X_\mu{}^\alpha{}_\beta)^\dagger=X_\mu{}_\beta{}^\alpha$.
We henceforth fix $\xi$ in the canonical gauge $\xi=\eta$,  
where the theory remains invariant only under the local transformation $S$ satisfying (\ref{SGT}). 

The gauge invariant Lagrangian for the gauge field $X_\mu$ is given by 
\begin{eqnarray}
{\cal L}_X&=&-\frac{1}{4}{\rm Tr} X^{\mu\nu}X_{\mu\nu}
=-\frac{1}{4}X^{\mu\nu\alpha}{}_\beta X_{\mu\nu}{}^\beta{}_\alpha,\\
X_{\mu\nu}{}^\alpha{}_\beta
&=&\partial_\mu X_\nu{}^\alpha{}_\beta-\partial_\nu X_\mu{}^\alpha{}_\beta
+ig(X_\mu{}^\alpha{}_\gamma X_\nu{}^\gamma{}_\beta-X_\nu{}^\alpha{}_\gamma X_\mu{}^\gamma{}_\beta).
\end{eqnarray}

In order to quantize the theory in the interaction representation, 
we have to add to the Lagrangian the gauge fixing term 
${\cal L}_{\rm gf}$, and the Faddeev-Popov ghost term ${\cal L}_{\rm gh}$\cite{FP}:
\begin{eqnarray}
{\cal L}_{\rm gf}&=&-\frac{1}{2}{\rm Tr}(\partial\cdot X)^2=-\frac{1}{2}\partial^\mu X_\mu{}^\alpha{}_\beta\partial^\nu X_\nu{}^\beta{}_\alpha,\\
{\cal L}_{\rm gh}&=&{\rm Tr}\partial^\mu\bar{\theta} D_\mu\theta
={\rm Tr}\partial^\mu\bar{\theta}(\partial_\mu\theta+ig[X_\mu, \theta])
=\partial^\mu\theta^{\dagger\alpha}{}_\beta[\partial_\mu\theta^\beta{}_\alpha+ig(X_\mu{}^\beta{}_\gamma\theta^\gamma{}_\alpha-\theta^\beta{}_\gamma X_\mu{}^\gamma{}_\alpha)],
\end{eqnarray}
where $\bar{\theta}=\eta\theta^\dagger\eta\neq\theta$. 
Further, we have to impose the physical condition by Fermi \cite{Fermi}, or by Gupta-Bleuler \cite{Gupta,Bleuler}, 
on the state vector:
\begin{equation}
\partial\cdot X^\alpha{}_\beta\vert\Phi\rangle=0, \quad {\rm or} \quad 
\langle\bar{\Phi}\vert\partial\cdot X^\alpha{}_\beta\vert\Phi\rangle=0. 
\end{equation}
Dividing the total Lagrangian ${\cal L}={\cal L}_\chi+{\cal L}_X+{\cal L}_{\rm gf}+{\cal L}_{\rm gh}$ 
into the free part ${\cal L}_0$, and the interaction part ${\cal L}_{\rm int}={\cal L}-{\cal L}_0$, 
where 
\begin{equation}
{\cal L}_0=\chi_\alpha^\dagger\bar{\sigma}^\mu{i}\partial_\mu\chi^\alpha-\frac{1}{2}\partial^\mu X^\nu{}^\alpha{}_\beta\partial_\mu X_\nu{}^\beta{}_\alpha
+\partial^\mu\theta^{\dagger\alpha}{}_\beta\partial_\mu\theta^\beta{}_\alpha,
\end{equation}
we obtain from ${\cal L}_0$ the non-zero canonical equal time (anti-)commutation relations
\begin{eqnarray}
\{\chi^\alpha(t,\bm{x}),\chi_\beta^\dagger(t,\bm{x}')\}&=&\xi^\alpha{}_\beta\delta^3(\bm{x}-\bm{x}'),\\ 
\ [X_\mu{}^\alpha{}_\beta(t,\bm{x}),\dot{X}^\nu{}^\gamma{}_\delta(t,\bm{x}')]
 &=&-i\eta_\mu{}^\nu\xi^\alpha{}_\delta\xi_\beta{}^\gamma\delta^3(\bm{x}-\bm{x}'),\\
\{\theta^\alpha{}_\beta(t,\bm{x}), \dot{\theta}^{\dagger\gamma}{}_\delta(t,\bm{x}')\}
&=&i\xi^\alpha{}_\delta\xi_\beta{}^\gamma\delta^3(\bm{x}-\bm{x}').
\label{CRX}
\end{eqnarray}
We define as before the reciprocal conjugate operator $\eta$ by 
\begin{equation}
\eta\chi^\alpha\eta=\chi_\alpha, \quad \eta{X}_\mu{}^\alpha{}_\beta\eta=X_{\mu\alpha}{}^\beta, \quad
\eta\theta^\alpha{}_\beta\eta=\theta_\alpha{}^\beta.
\label{RCO}
\end{equation}
Since the total Hamiltonian density is
\begin{equation}
{\cal H}=i\chi_\alpha^\dagger\dot{\chi}^\alpha-\dot{X}^{\mu\beta}{}_\alpha\dot{X}_\mu{}^\alpha{}_\beta
+\dot{\theta}^{\dagger\beta}{}_\alpha\dot{\theta}^\alpha{}_\beta-{\cal L}={\cal H}_0+{\cal H}_{\rm int},
\end{equation}
we have
\begin{equation}
[\eta, {\cal L}_0]=[\eta, {\cal L}_{\rm int}]=[\eta, {\cal H}_0]=[\eta, {\cal H}_{\rm int}]=0.
\label{CC}
\end{equation}
The gauge fixing terms ${\cal L}_{\rm gf}$ and ${\cal L}_{\rm gh}$ do not affect the reciprocal conjugate invariance of the partial Lagrangians and the partial Hamiltonians.

The Schr\"{o}dinger equation in the interaction representation is now
\begin{equation}
i\frac{d}{dt}\vert\Phi\rangle={H}_{\rm int}\vert\Phi\rangle, \quad {H}_{\rm int}=-\int d^3x{\cal L}_{\rm int},
\end{equation}
and the probability: $\langle\bar{\Phi}\vert\Phi\rangle=\langle\Phi\vert\eta\vert\Phi\rangle$ conserves:
\begin{equation}
\frac{d}{dt}\langle\bar{\Phi}\vert\Phi\rangle=i\langle\Phi\vert[H_{\rm int}, \eta]\vert\Phi\rangle=0.
\end{equation}

For confirming the special gauge invariance of the probability, we have to consider the Fock state $\vert\Phi\rangle$ composed only of the physical particles, excluding ghosts. 
Though the total Lagrangian is not locally invariant after the gauge fixing, 
the global gauge invariance is retained for both ${\cal L}_0$ and ${\cal L}_{\rm int}$. 
Then, we obtain the operations of the global transformation $S$ on $\chi^\alpha$,  $\chi^\dagger_\alpha$, and $X_\mu{}^\alpha{}_\beta$ as 
\begin{equation}
{\chi'}^\alpha=S\chi^\alpha{S}^{-1}, \quad 
{\chi'}^\dagger_\alpha=S\chi^\dagger_\alpha{S}^{-1}, \quad 
X'_\mu{}^\alpha{}_\beta=SX_\mu{}^\alpha{}_\beta S^{-1},
\label{ST1}
\end{equation}
where the second equality is derived from (\ref{SGT}) and (\ref{RCO}). 
Since the Fock state generally has the form:
\begin{equation}
\vert\Phi\rangle=\prod_A(\int{A}_\alpha\chi^\alpha)\prod_B(\int{B}^\beta\chi^\dagger_\beta) \prod_C(\int{C}^\mu{}_\gamma{}^\delta X_\mu{}^\gamma{}_\delta)\rangle,
\end{equation}
where $A$, $B$, and $C$ are spacetime functions, we have
\begin{equation}
\vert\Phi'\rangle=S\vert\Phi\rangle, \quad \langle\bar{\Phi}'\vert=\langle\bar{\Phi}\vert{S}^{-1},
\end{equation}
and therefore the probability $\langle\bar{\Phi}\vert\Phi\rangle$ is globally gauge invariant. 

In order to verify the positivity of the conserved invariant probability, 
it is suffice to consider only the physical particles in 4-momentum and helicity eigenstates.  
The field operators $\chi^\alpha(x)$ and $X_\mu{}^\alpha{}_\beta(x)$ in the interaction representation expanded in terms of the creation and annihilation operators are
\begin{eqnarray}
\chi^\alpha(x)&=&\frac{1}{\sqrt{V}}\sum_{\bm{p}} L_{\bm{p}}
\left(a_{\bm{p}}{}^\alpha{e}^{-ip\cdot x}+b^\dagger_{\bm{p}}{}^\alpha{e}^{ip\cdot x}\right), 
\quad 
\{a_{\bm{p}}{}^\alpha,a_{\bm{p}'\beta}^\dagger\}
=\{b_{\bm{p}\beta},b^\dagger_{\bm{p}'}{}^\alpha\}
=\xi^\alpha{}_\beta\delta^3_{\bm{p}\bm{p}'},\\
X_\mu{}^\alpha{}_\beta(x)
&=&\sum_{\bm{k}}\frac{1}{\sqrt{2kV}}
\left(c_{\bm{k}\mu}{}^\alpha{}_\beta e^{-ik\cdot x}+c^\dagger_{\bm{k}\mu}{}^\alpha{}_\beta e^{ik\cdot x}\right), \quad
\left[c_{\bm{k}\mu}{}^\alpha{}_\beta,c^\dagger_{\bm{k}'\nu}{}^\gamma{}_\delta\right]
=-\eta_{\mu\nu}\xi^\alpha{}_\delta\xi_\beta{}^\gamma\delta^3_{\bm{k}\bm{k}'}, \\
\end{eqnarray}
where $L_{\bm{p}}$ is the left-handed helicity eigenfunction satisfying 
$\bm{\sigma}\cdot\bm{p}L_{\bm{p}}=-\vert{\bm{p}}\vert L_{\bm{p}}$.  
The physical one-particle states are therefore expressed as 
$\vert\Phi_{\rm f}\rangle=a^\dagger_{\bm{p}\alpha}\rangle$ for of a fermion, 
$\vert\Phi_{\rm f}\rangle=b^\dagger_{\bm{p}}{}^\alpha\rangle$ for an anti-fermion, 
and $\vert\Phi_{\rm g}\rangle=c^\dagger_{\bm{k}s}{}^\alpha{}_\beta\rangle$ for a gauge boson, where 
\begin{equation}
c_{\bm{k}s}{}^\alpha{}_\beta:=\epsilon^\mu{}_s c_{\bm{k}\mu}{}^\alpha{}_\beta, \quad 
k\cdot\epsilon_s=0, \quad
\epsilon_s^*\cdot\epsilon_{s'}=-\delta_{ss'}, \quad s=\pm1,
\label{QCV}
\end{equation}
and $\epsilon^\mu{}_s$'s are the circular polarization vectors. 
Further from (\ref{RCO}), the reciprocal conjugate transformations for the creation and annihilation operators are  
\begin{equation}
\eta a_{\bm{p}}{}^\alpha\eta=a_{\bm{p}\alpha}, \quad \eta b_{\bm{p}}{}^\alpha\eta=b_{\bm{p}\alpha}, \quad 
\eta c_{\bm{k}\mu}{}^\alpha{}_\beta\eta=c_{\bm{k}\mu\alpha}{}^\beta,  
\label{RC}
\end{equation}
while the free Hamiltonians for the fermions and the gauge bosons are
\begin{equation}
H^0_\chi=\sum_{\bm{p}}\vert\bm{p}\vert\left(a_{\bm{p}\alpha}^\dagger a_{\bm{p}}{}^\alpha
+b^\dagger_{\bm{p}}{}^\alpha b_{\bm{p}\alpha}-d\right), \quad 
H^0_X=\sum_{\bm{k}}\vert\bm{k}\vert\left(-c^{\dagger\mu\alpha}_{\bm{k}}{}_\beta c_{\bm{k}\mu}{}^\beta{}_\alpha
+d^2\right),
\label{FreeHamiltonianForX}
\end{equation}
where the zero-point energy for the unphysical modes are omitted. 

Taking care of conventions: $(c_{\bm{k}\mu}{}^\alpha{}_\beta)^\dagger=c^\dagger_{\bm{k}\mu}{}_\beta{}^\alpha$, 
and 
$(c^\dagger_{\bm{k}\mu}{}^\alpha{}_\beta)^\dagger=c_{\bm{k}\mu\beta}{}^\alpha$, we have 
\begin{eqnarray}
H_\chi\vert\Phi_{\rm f}\rangle=\vert\bm{p}\vert\vert\Phi_{\rm f}\rangle,
\quad
\langle\bar{\Phi}_{\rm f}\vert H_\chi\vert\Phi_{\rm f}\rangle=\vert\bm{p}\vert, \\
H_X\vert\Phi_{\rm g}\rangle=\vert\bm{k}\vert\vert\Phi_{\rm g}\rangle,
\quad
\langle\bar{\Phi}_{\rm g}\vert H_X\vert\Phi_{\rm g}\rangle=\vert\bm{k}\vert,
\end{eqnarray}
which guarantee that the invariant probability is generally positive definite even for many particle states, 
irrespective of the $\eta$-parity of each particle.  
No negative probabilities appear also in the Pauli's sense \cite{Pauli}, since the energy eigenvalue and the expectation value of the energy is the same.   

In the canonical quantum field theory, the scattering amplitudes and the effective action are calculated from the vacuum expectation value of some multiple products of the interaction Lagrangian, where the vacuum state is assumed to be gauge invariant. 
The gauge indices in vacuum amplitudes come from the gauge factors in the time-ordered two-point Green's functions: 
\begin{eqnarray}
\langle{\rm T}\chi^\alpha(x)\chi_\beta^\dagger(x')\rangle&=&\xi^\alpha{}_\beta S_F(x-x'), \\
\langle{\rm T}X_\mu{}^\alpha{}_\beta(x)X_\nu{}^\gamma{}_\delta(x')\rangle&=&-\eta_{\mu\nu}
\xi^\alpha{}_\delta\xi_\beta{}^\gamma D_F(x-x'),\\
\langle{\rm T}\theta^\alpha{}_\beta(x)\theta^{\dagger\gamma}{}_\delta(x')\rangle&=&
\xi^\alpha{}_\delta\xi_\beta{}^\gamma D_F(x-x'),
\end{eqnarray}
where $S_F(x-x')$ and $D_F(x-x')$ are the time-ordered two point functions for a Weyl fermion and a massless scalar boson, respectively.
However, the final results are invariant under special gauge transformations, where all the gauge indices are contracted by the rule $\xi^\alpha{}_\alpha=d$, and no dependence on the signature of the metric remains. 
This will be the case also for the masses and the charges of particles, the renormalization constants, and the renormalization group equations. 
Accordingly, the physical results obtained from the general gauge theory of $d$-dimensions will be the same as those obtained from the SU($d$)$\times$U(1) gauge theory, irrespective of the signature of the internal metric. We may call the phenomenon ``the law of inertia" in general gauge theory, in analogy with the Sylvester's law of inertia.
\section{Spontaneous fusion of overlapping Lorentz symmetries}
As the spacetime symmetry breaking mechanism consistent with relativity, we consider the fusion of two Lorentz symmetries, one is in the gauge space, the other in the real Minkowski spacetime.

The motivation of thinking spacetime symmetry breakdown seems, on the theoretical side, 
due to the desire to understand the existence of massless particles appearing in nature, neutrinos, photons, and possibly gravitons, as the Nambu-Goldstone modes \cite{NJ,Goldstone,GSW} emergent from broken symmetries 
\cite{Bjorken,Eguchi,BK}. 
Then, the Goldstone particles with spins necessarily require the breakdown of spacetime symmetry, 
or the Lorentz violation. 

Another motivation is to understand the disparity in the electroweak interactions between the left-handed doublet and the right-handed singlet \cite{KN1,KN2,KN3}. 
It would be natural, if the right-handed fermions are emergent ones from the left-handed multiplets by the violation of Lorentz invariance. Further, the baryon asymmetry, or the absence of antimatter in the universe 
may also become easier  to understand, if its origin is in the spontaneous violation of Lorentz invariance. 

In those cases, however, it is also required that these particles should have relativistic kinematics and interactions, which is in a dilemma. 
Is there any mechanism of symmetry breaking in which the same symmetry survives as that broken? 
One of the answer will be the spontaneous symmetry fusion.

The generalized gauge theory allows Lorentz symmetry as the subgroup of gauge symmetry, 
and then the model on this basis can have a double Lorentz symmetry. 
If symmetry breaking occurs in such a way that the two symmetries are glued together 
as the result of the vacuum expectation values developed by some mixed tensor fields, 
the double symmetry reduces to one, and the combined Lorentz symmetry will survive in the resultant theory, as if it were not broken.

For example, a Lorentz invariant SU($N$)$\times$U(1) gauge theory will be obtained from spontaneous spacetime symmetry fusion, when a general gauge theory has the Minkowski metric of $N+4$ dimensions in the gauge space.
In this case, $N+4$ dimensional complex Lorentz group QL($N$+4) breaks down to QL(4)$\times$SU($N$), and  
then, QL(4) fuses with the real spacetime Lorentz group SL(4) into SL(4)$\times$U(1),
though these two events actually happen at the same time. 
Eventually, we have SL(4)$\times$QL($N$+4)$\rightarrow$SL(4)$\times$SU($N$)$\times$U(1). 
The desired symmetry breaking will be realized, 
when the vacuum expectation value of the gauge potential $X_{\mu AB}$ takes the block off-diagonal form:
\begin{equation}
\langle X_{\mu\alpha\beta}\rangle=0, \quad 
\langle X_{\mu\alpha a}\rangle=\langle X^*_{\mu a\alpha}\rangle\simeq\langle \phi_a\rangle\eta_{\mu\alpha}\neq0, 
\quad \langle X_{\mu ab}\rangle=0,
\end{equation}
where $\alpha$, and $\beta$ are the internal Lorentz indices, while $a$ and $b$ are SU($N$) indices. 

However, a possible problem arising here is the violation of the condition (\ref{CC}), 
which is required for the conservation of the invariant positive probability. 
In fact, we see that the Hamiltonian $H^0_\chi$ for the free motion of the chiral multiplet $\chi$ does not commute with $\eta$ after symmetry breaking:
\begin{equation}
i\dot{\chi}=H^0_\chi\chi, \quad H^0_\chi=-\bm{\sigma}\cdot\bm{p}+\langle \bar{\sigma}^\mu X_\mu\rangle, \quad 
[\eta, \langle \bar{\sigma}^\mu X_\mu\rangle]\neq0.
\end{equation}
Then, the positive definite probability will not conserve, or the indefinite probability revives. 
Does the symmetry breakdown also break the consistency of the general gauge theory? 

While a consistent quantum theory will not be hoped for the whole quantum states, 
we may expect the existence of the subdomain in the space of quantum states, 
in which quantum physics remains consistent,  
and the particles and the interactions are closed in the subdomain with the conserving positive definite probabilities. 
Such a subdomain may be extracted by imposing the consistency conditions:
\begin{equation}
\langle\bar{\Phi}\vert[\eta, {\cal L}]\vert\Phi\rangle=0, \quad \langle\bar{\Phi}\vert\Phi\rangle>0.
\end{equation}

The extraction can be actually performed by parametrizing the field variables in terms of those conventionally used in the ordinary quantum field theory, so as to eliminate broken internal Lorentz indices, 
with taking care of the Lorentz invariance of the final results. 
How to find the parametrization  can be learned from the solutions of the free equations of motion, 
as shown in the next section.  
\section{A chiral sextet model of leptons}
We take 6 dimensional complex Lorentz group as the gauge symmetry, and index its vector representation  by $A=(\alpha,a)$, where 
$\alpha=0,1,2,3$, and $a=1,2$.
The internal metric is $\xi_{AB}=(\eta_{\alpha\beta}, -\delta_{ab})$.
The Lagrangian in four dimensional Minkowski spacetime is then given by
\begin{eqnarray}
{\cal L}&=&{\cal L}_\chi+{\cal L}_X,\\
{\cal L}_\chi&=&-\chi_A^\dagger\bar{\sigma}\cdot p\chi^A+\frac{g}{\sqrt{2}}X_{\mu AB}\chi^{A\dagger}\bar{\sigma}^\mu\chi^B, \label{LaForChi}\\
{\cal L}_X&=&-\frac{1}{4}X^{\mu\nu AB}X_{\mu\nu BA},\\
X_{\mu\nu AB}&:=&\partial_\mu X_{\nu AB}-\partial_\nu X_{\mu AB}
+i\frac{g}{\sqrt{2}}(X_{\mu A}{}^CX_{\nu CB}-X_{\nu A}{}^CX_{\mu CB}),
\end{eqnarray}
where we have changed the coupling constant from $g$ to $g/\sqrt{2}$, which makes the new $g$ identifiable with the conventional weak coupling constant.
The gauge fields in the components $X_{\mu AB}^*=X_{\mu BA}$ include the U(1) potential with the same coupling constant $g$. 

We extract the physical domain by expressing $\chi^A$ and $X_{\mu AB}$ by
\begin{eqnarray}
\chi^\alpha&=&\frac{1}{\sqrt{2}}\left(\sigma^\alpha e_R-\frac{2p^\alpha}{m}e_L\right),\quad 
\chi^{3+a}=L^a=\left(\begin{array}{c}\nu_L\\ e_L \end{array}\right), \label{PramChi}\\
X_{\mu\alpha\beta}&=&Y_{\mu\alpha\beta}
=\frac{1}{\sqrt{6}}\left(\eta_{\mu\alpha}Y_\beta+\eta_{\mu\beta}Y_\alpha-\eta_{\alpha\beta}Y_\mu
-i\epsilon_{\mu\alpha\beta\nu}Y^\nu\right),\label{ParamX1}\\
X_{\mu a\alpha }&=&\frac{\phi_a}{\sqrt{3}}\eta_{\mu\alpha}, \quad X_{\mu\alpha a}=\frac{\phi^*_a}{\sqrt{3}}\eta_{\mu\alpha}, \label{ParamX2} \\
X_{\mu ab}&=&A_{\mu ab}-\frac{\xi_{ab}}{\sqrt{6}}Y_\mu, \quad A_\mu{}^a{}_a=0,
\label{ParamX3}
\end{eqnarray}
where $\nu_L$, $e_L$, and $e_R$ satisfy the equations of motion:
\begin{equation}
\bar{\sigma}\cdot p \nu_L=0, \quad \bar{\sigma}\cdot pe_L=me_R, \quad \sigma\cdot pe_R=me_L.
\label{EqOfMotionForLeptons}
\end{equation}
The mass $m$ for $e_L$ and $e_R$ will be determined later.

It is convenient to represent SU(2) part of the scalar and  the gauge fields in the matrix notation:
\begin{equation}
\phi^a=-\phi_a=(\bm{\phi})^a,\quad 
A_\mu{}^a{}_b=-A_{\mu ab}=\left(\bm{\rho}\cdot\bm{A}_\mu/\sqrt{2}\right)^a_{\ b},
\end{equation}
where $\bm{\rho}/2$ are the SU(2) generators. 
Then, the Lagrangian ${\cal L}_\chi$ reduces to
\begin{equation}
{\cal L}_\chi=e_R^\dagger\sigma^\mu i\left(\partial_\mu-\frac{ig}{\sqrt{3}}Y_\mu\right)e_R
+L^\dagger\bar{\sigma}^\mu 
i\left(\partial_\mu+\frac{ig}{2}\bm{\rho}\cdot\bm{A}_\mu-\frac{ig}{2\sqrt{3}}Y_\mu\right)L
-\frac{g}{\sqrt{3}}\left(L^\dagger\bm{\phi}e_R+e_R^\dagger\bm{\phi}^\dagger L\right),
\label{Lag6}
\end{equation}
where the following identities have been used: 
$Y_{\mu\alpha\beta}\chi^{\alpha\dagger}\bar{\sigma}^\mu\chi^\beta
=(Y_\mu/\sqrt{6})\chi^{\alpha\dagger}\bar{\sigma}_\alpha\sigma^\mu\bar{\sigma}_\beta\chi^\beta$, and $\bar{\sigma}_\alpha\chi^\alpha=\sqrt{2}e_R$.

The expression (\ref{Lag6}) should be understood as the interaction representation with respect to the spinor fields, since we have used the equations of motion (\ref{EqOfMotionForLeptons}) 
for the derivation. Therefore, the free part of the Lagrangian is actually zero. 

The Lagrangian ${\cal L}_X$ in the physical domain is, on the other hand, calculated as
\begin{eqnarray}
{\cal L}_X&=&
-\frac{1}{4}Y^{\mu\nu}Y_{\mu\nu}
-\frac{1}{4}\bm{A}^{\mu\nu}\cdot\bm{A}_{\mu\nu}
+D^\mu\bm{\phi}^\dagger D_\mu\bm{\phi}-\frac{g^2}{3}(\bm{\phi}^\dagger\bm{\phi})^2,\\
D_\mu\bm{\phi}&=&\left(\partial_\mu+\frac{ig}{2}\bm{\rho}\cdot\bm{A}_\mu+\frac{ig}{2\sqrt{3}}Y_\mu\right)\bm{\phi}.
\end{eqnarray}

The final form of the total Lagrangian ${\cal L}={\cal L}_\chi+{\cal L}_X$ is Lorentz invariant, and has a positive definite metric in the gauge space, despite starting from a theory with an indefinite metric with broken Lorentz invariance. 
The result is almost identical with the standard electroweak theory before the SU(2)$_{\rm L}\times$U(1) symmetry breaking. 
The only difference is the absence of symmetry breaking term quadratic with respect to the scalar doublet. 
The missing term may be obtained, though quadratically divergent, by the fermion loop collection, 
since there is no Ward identity for scalar fields.

We remark that the right-handed electron emerges from the left-handed quartet $\chi^\alpha$.   
If $\bm{\phi}$ is identified with the Higgs doublet\cite{Higgs,Englert}, 
the Higgs coupling to the electron is also correctly derived as in the standard electroweak theory\cite{Weinberg}. 

The hyper U(1) coupling constant $g'$ in the standard theory can be read $g/\sqrt{3}$, 
which predicts the Weinberg angle $\sin^2\theta_W=1/4$. 
The agreement with the observed value $\sin^2\theta_W\simeq0.23$ seems fairly good.

Further, the hyper charges, $1$,$-1$, and $-2$ for the Higgs doublet $\bm{\phi}$,  
the left-handed doublet $L$ and the right-handed singlet $e_R$, respectively, are completely identical with that assigned in the standard electroweak theory. 

Another prediction concerns the coupling constant $\lambda$ for the quartic term in the Higgs potential: $-\lambda(\bm{\phi}^\dagger\bm{\phi})^2$, for which our theory gives $g^2/3$. 
If we put $\lambda=g^2/n$ in the standard model, then, $n=8(m_W/m_H)^2\simeq3.3$ 
for $m_W\sim80$GeV and $m_H\sim125$GeV, which is comparable enough to our result.

The masses of the Higgs boson and the weak bosons are obtained by 
adding the term $(2g^2/3)\phi_0^2\bm{\phi}^\dagger\bm{\phi}$ with some constant  
to the Higgs potential. 
Then, 
\begin{equation}
\bm{\phi}=\left(\begin{array}{c}0\\\phi_0+\sigma/\sqrt{2} \end{array}
\right), \quad 
Z_\mu:=\frac{\sqrt{3}}{2}A_{\mu3}-\frac{Y_\mu}{2}, \quad 
A_\mu:=\frac{A_{\mu3}}{2}+\frac{\sqrt{3}}{2}Y_\mu, \quad
W_\mu:=\frac{A_{\mu1}+iA_{\mu2}}{\sqrt{2}},
\end{equation}
\begin{eqnarray}
D^\mu\bm{\phi}^\dagger D_\mu\bm{\phi}-\frac{g^2}{3}(\bm{\phi}^\dagger\bm{\phi}-\phi_0^2)^2
=&&\frac{1}{2}\partial^\mu\sigma\partial_\mu\sigma-\frac{2}{3}g^2\phi_0^2\sigma^2
-\frac{\sqrt{2}}{3}g^2\phi_0\sigma^3-\frac{g^2}{12}\phi_0^2\sigma^4 \nonumber\\
&&+\frac{g^2\phi_0^2}{2}(1+\frac{\sigma}{\sqrt{2}\phi_0})^2(W^{\mu\dagger}W_\mu+\frac{2}{3}Z^\mu Z_\mu),
\end{eqnarray}
and we have
\begin{equation}
m_Z=\sqrt{\frac{2}{3}}g\phi_0, \quad m_W=\frac{\sqrt{3}}{2}m_Z, \quad  m_H=\sqrt{2}m_Z.
\end{equation}
The last equation predicts the Higgs boson mass from the $Z$-boson mass, which is $m_H\sim127$GeV for $m_Z\sim90$GeV. 
The interaction terms for the leptons in the Lagrangian ${\cal L}_\chi$ (\ref{Lag6}) 
expressed in terms of Dirac spinors are 
\begin{eqnarray}
{\cal L}_\chi^{\rm int}&=&\frac{g}{2}A_\mu\bar{e}\gamma^\mu{e}
+\frac{g}{2\sqrt{3}}Z_\mu\left[\bar{e}\gamma^\mu\gamma_5{e}-\bar{\nu}\gamma^\mu(1-\gamma_5)\nu\right]\nonumber\\
&&-\frac{g}{2\sqrt{2}}W_\mu\bar{e}\gamma^\mu(1-\gamma_5)\nu
-\frac{g}{2\sqrt{2}}W^\dagger_\mu\bar{\nu}\gamma^\mu(1-\gamma_5)e\label{InterractionForLeptons}\\
&&-\frac{g\phi_0}{\sqrt{3}}(1+\frac{\sigma}{\sqrt{2}\phi_0})\bar{e}e\nonumber.
\end{eqnarray}
We see that the unit of the electric charge is $e=g/2=g\sin\theta_W$, as expressed in the standard theory. 
The electron has the pure axial vector coupling to the $Z$-boson. 

It is important to notice that a chiral gauge theory has derived the vector coupling of the electromagnetic interaction and the scalar coupling of the Higgs interaction, 
since the result suggests that the parity invariance may not be an intrinsic property, but an acquired characteristic in the laws of nature.

As for the fermion mass, on the other hand, we see from Eq.(\ref{InterractionForLeptons}) that $m=g\phi_0/\sqrt{3}=m_Z/\sqrt{2}$, which is too big for the electron, or even for the tau lepton. However, as the mass for the weak doublet including quarks, it ranges from $0$ to the top quark mass: $m_{\rm top}\simeq174$GeV.

Since the energy scale of our model is determined only by the missing symmetry breaking term, every mass scale can not be different from that of the weak bosons or the Higgs boson. The mass spectrum of the fundamental fermions is even now total mystery for us, and no theory, including our model, seems to provide the mechanism for the mass quantization. 

Further, several remarks are in order.
We have seen that generalized gauge theory and spontaneous spacetime symmetry fusion provides an internal unification of the electroweak interactions, 
in which the photon, the weak bosons, and the Higgs boson are in the same multiplet $X_{\mu AB}$, 
while the neutrino, the left-handed ``electron", and the right-handed ``electron" are coming from the same left-handed sextet $\chi^A$. 
There are no need to take charge conjugation to 
include the right-handed fermions in a chiral multiplet, as in the SU(5) model \cite{GG}. 
The emergence of a right-handed fermion from the left-handed multiplet will be a characteristic phenomenon of Lorentz violation. Despite that, there still remains Lorentz invariance in the resultant theory. 
Accordingly, the spontaneous fusion of double Lorentz symmetry solves the dilemma between spacetime symmetry breakdown and Lorentz invariance. 
\section{Lorentz quartet}
Our leptons come from the sextet of the left-handed Weyl spinor, consisting of the Lorentz quartet $\chi^\alpha$, and the SU(2) doublet $\chi^a=(\chi^4,\chi^5)$. Of those, one of the quartet becomes the right-handed ``electron" $e_R$ after spacetime symmetry breaking. Then, what about the remaining triplet?  

The free equations of motion for $\chi^A$ after symmetry breaking is read from Eq.(\ref{LaForChi}) as
\begin{eqnarray}
\bar{\sigma}\cdot{p}\chi^\alpha&=&-\frac{m}{\sqrt{2}}\bar{\sigma}^\alpha\chi^5,\\
\bar{\sigma}\cdot{p}\chi^4&=&0,\\
\bar{\sigma}\cdot{p}\chi^5&=&\frac{m}{\sqrt{2}}\bar{\sigma}_\alpha\chi^\alpha,\\
\end{eqnarray}
where we have put $m=g\phi_0/\sqrt{3}$.
These equations have one solution for $p^2=m^2$:
\begin{equation}
\chi^\alpha=\frac{1}{\sqrt{2}}\left(\sigma^\alpha e_R-\frac{2p^\alpha}{m}e_L\right), \quad \chi^4=0, \quad \chi^5=e_L,
\end{equation}
and the three solutions for $p^2=0$: 
\begin{eqnarray}
\nu^\alpha_1&=&\frac{p^\alpha}{m}\nu_L, \quad \nu^4_1=0, \quad \nu^5_1=0,\\
\nu^\alpha_2&=&\left(\bar{\sigma}^\alpha+2\eta^{\alpha0}-2itp^\alpha\right)\nu_L, \quad 
\nu^4_2=0, \quad \nu^5_2=-\frac{\sqrt{8}p^0}{m}\nu_L,\\
\nu^\alpha_3&=&e_-^\alpha\nu_L, \quad \nu^4_3=0, \quad \nu^5_3=0,
\end{eqnarray}
where $\bar{\sigma}\cdot{p}\nu_L=0$, and $e_-^\alpha$ is the left-circular polarization 4-vector, perpendicular to the 4-momentum: $p\cdot e_-=0$. 
These solutions $\nu_1$, $\nu_2$, and $\nu_3$ are therefore neutrinos. 

The $\nu_1$ neutrino has zero norm, while that of the resonant neutrino $\nu_2$ is indefinite, depending on its momentum. The two neutrinos are not orthogonal with each other. 
Then, by taking appropriate linear combinations of the two,  
we can make them a pair of zero-norm neutrinos to exclude from the physical domain as undetectable. 
The remaining $\nu_3$ has a positive norm, and can remain in the physical domain, 
though it has a weird relativistic property. However, this extra neutrino decouples from the electroweak interactions.
Actually the $\nu_3$ neutrino has the electroweak interactions only through the hyper current:
\begin{equation}
Y_{\mu\alpha\beta}\nu_3^{\alpha\dagger}\bar{\sigma}^\mu\nu_3^\beta
=\frac{Y_\mu}{\sqrt{6}}(\bar{\sigma}\cdot{e}_-\nu_L)^\dagger\sigma^\mu(\bar{\sigma}\cdot{e}_-\nu_L), 
\end{equation}
which vanishes, since $\bar{\sigma}\cdot{e}_-\nu_L=0$ for $\bar{\sigma}\cdot{p}\nu_L=0$. 
\section{Electroweak Gravity}
We show further that our model generates also gravitational interactions, 
if spontaneous spacetime symmetry fusion occurs at a local frame in curved spacetime. 
In this case, the index of the quantum Lorentz group ``$\alpha$" in ``$A$" is identified with the local Lorentz frame index 
$\alpha$ after symmetry fusion.  This implies that the Weyl quartet $\chi^\alpha$ turns into a chiral vector-spinor field with spin $3/2$ after symmetry fusion. 
Then, the local Lorentz invariance requires the extra local Lorentz connection in the covariant derivative for $\chi^\alpha$. 
Therefore the parametrization of the gauge field $X_{\mu AB}$ in Eq.(\ref{ParamX1}), Eq.(\ref{ParamX2}), and Eq.(\ref{ParamX3}) should take the form:
\begin{eqnarray}
X_{\mu\alpha\beta}&=&Y_{\mu\alpha\beta}-i\frac{\sqrt{2}}{g}\omega_{\mu\alpha\beta},\\
Y_{\mu\alpha\beta}&=&\frac{1}{\sqrt{6}}\left(e_{\mu\alpha}Y_\beta+e_{\mu\beta}Y_\alpha-\eta_{\alpha\beta}Y_\mu-i\epsilon_{\mu\alpha\beta\nu}Y^\nu\right),\label{Y_mu_alpha_beta}\\
X_{\mu a\alpha }&=&\frac{\phi_a}{\sqrt{3}}e_{\mu\alpha}, \quad X_{\mu\alpha a}=\frac{\phi^*_a}{\sqrt{3}}e_{\mu\alpha}, \label{GX_a_alpha}\\
X_{\mu ab}&=&A_{\mu ab}-\frac{\xi_{ab}}{\sqrt{6}}Y_\mu, \quad A_\mu{}^a{}_a=0,
\end{eqnarray}
where $e_{\mu\alpha}$ is the vierbein invariant under the parallel transportation:
\begin{equation}
\nabla_\rho e_{\mu\alpha}
=\partial_\rho e_{\mu\alpha}-\Gamma^\nu_{\rho\mu}e_{\nu\alpha}+\omega_{\rho\alpha}{}^\beta e_{\mu\beta}=0.
\end{equation}
This condition is necessary for a conserved current to be conserved also in a local Lorentz frame: 
$\nabla_\mu J^\mu=\nabla_\alpha J^\alpha=0$. 

For the Weyl sextet, we notice that the 4-momentum operator $p^\alpha$ in the parametrization (\ref{PramChi})
should be the covariant derivative. 
Since the covariant momenta do not commute: 
\begin{equation}
[p_\mu,p_\nu]e_L=-\frac{1}{8}R_{\mu\nu\alpha\beta}\bar{\sigma}^{\alpha\beta}e_L,
\end{equation}
where 
$R_{\mu\nu\alpha\beta}=R_{\mu\nu\rho\sigma}e^\rho{}_\alpha e^\sigma{}_\beta$ is the Riemann curvature,  
the following extra terms appear in the sextet Lagrangian (\ref{Lag6}) in curved spacetime:
\begin{equation}
-\frac{R}{4m}(e_L^\dagger e_R+e_R^\dagger e_L)
+\frac{R_{\alpha\beta}}{m^2}e_L^\dagger\bar{\sigma}^\alpha p^\beta e_L.
\label{LLagInCSP}
\end{equation}
The first term is the Higgs-like coupling of the scalar curvature to the ``electron", 
while the second term can be viewed as the gravitational correction to the vierbein:
\begin{equation}
e_{\mu\alpha}\rightarrow e_{\mu\alpha}+\frac{R_{\mu\alpha}}{m^2},
\end{equation}
though only for the kinetic term of the left-handed ``electron". It shows therefore the gravitational parity violation. 

On the other hand, the final form of the Lagrangian for the gauge fields ${\cal L}_X$ is  
\begin{eqnarray}
{\cal L}_X&=&
D^\mu\bm{\phi}^\dagger D_\mu\bm{\phi}-\frac{g^2}{3}(\bm{\phi}^\dagger\bm{\phi})^2-\frac{1}{4}Y^{\mu\nu}Y_{\mu\nu}-\frac{1}{4}\bm{A}^{\mu\nu}\cdot\bm{A}_{\mu\nu} \nonumber\\
&&
-\frac{R}{3}\bm{\phi}^\dagger\bm{\phi}
-\frac{1}{2g^2}R^{\mu\nu\alpha\beta}R_{\mu\nu\alpha\beta},
+R_{\mu\nu}Y^\mu Y^\nu,
\label{LxInCSP}
\end{eqnarray}
where we have used the identity holding under the invariant spacetime integration:
\begin{equation}
\int d^4x\sqrt{-g}(\nabla_\mu Y^\mu)^2=\int d^4x\sqrt{-g}(\nabla^\mu Y^\nu\nabla_\nu Y_\mu+R_{\mu\nu}Y^\mu Y^\nu).
\end{equation}
The last three terms in (\ref{LxInCSP}) appear as the corrections in curved spacetime.
The first term couples the Higgs boson to the scalar curvature,  
and gives the Einstein action with the gravitational constant $G_{\rm N}=(3/16\pi)\phi_0^{-2}$, 
when the Higgs doublet develops the vacuum expectation value $(0,\phi_0$). 

The second term, which is originally the kinetic term for gauge fields before symmetry breaking, 
adds to the original Einstein equation the energy-momentum tensor for gravity itself, after the symmetry fusion. 

The third extra term is similar to the anomalous magnetic moment of the $W$-boson, and may be called the anomalous gravitational moment for both the photon and  the $Z$-boson, 
which would be absent, if $Y_\mu$ were really a U(1) gauge field.  
Actually, as seen from the parametrization (\ref{Y_mu_alpha_beta}), $Y_\mu$ is not a pure U(1) gauge field commuting with the other non-abelian components. 
The prediction and a comment on this term are found in \cite{BK}.

If $\phi_0$ is taken literary in the electroweak scale, we have
$G_{\rm N}=2\alpha/m_Z^2$, where $\alpha$ is the fine structure constant, which is too big for the Newton constant. Here appears again the discrepancy in the energy scale. 

Our theory is still devoid of the symmetry breaking term, which oughts to provide the energy scale for the theory. 
What is missing might be some quantum condition on the quadratic divergence in the quantum field theory, 
which could take various energy scales, including the Planck scale, in addition to the electroweak scale, in the Higgs potential. 

Even on such a hypothesis, our result may not exclude the gravitational interactions in the electroweak scale. If it exists, however, its effects should be restricted in a small spacetime volume compared with the astronomical scales, in order not to conflict with the celestial and cosmological phenomena. This hypothesis seems plausible, since the emergence of the gravitational interactions in the electroweak scale requires that spacetime symmetry fusion occurs at a local Lorentz frame in curved spacetime. 

Though our model lacks quarks and strong interactions, the quantum Lorentz group QL(6) includes SU(3) as the subgroup, though some part of which are overlapped with the hyper U(1) field and the local Lorentz connection. 
Wether it is possible for QL(6) to include QCD by adding some extra multiplets remains yet to be considered.

\end{document}